\documentclass[conference]{IEEEtran}

\usepackage{cite}
\usepackage{hyperref}
\usepackage{subcaption}
\usepackage[algo2e,ruled,vlined,linesnumbered]{algorithm2e}
\usepackage{amsmath}
\usepackage{amssymb}
\usepackage{mathtools}
\usepackage{dsfont}
\usepackage{physics}
\usepackage[capitalize,noabbrev]{cleveref}
\usepackage{orcidlink}
\usepackage[misc]{ifsym}
\usepackage{booktabs} 
\usepackage{multirow}
\usepackage{color}

\newcommand{\Cleafy}{Cleafy\xspace}
\newcommand{\cleafy}{Cleafy\xspace}
\newcommand{\osr}{\textsc{OSR}\xspace}
\newcommand{\gboost}{\textsc{Gradient Boosting}\xspace}
\newcommand{\maxlogit}{\textsc{MaxLogit}\xspace}
\newcommand{\logit}{\textsc{Logit}\xspace}
\newcommand{\logits}{\textsc{Logits}\xspace}
\newcommand{\osnn}{\textsc{OSNN}\xspace}
\newcommand{\fixnovel}{\textsc{Fixed-Novel}\xspace}
\newcommand{\loonovel}{\textsc{LOO-Novel}\xspace}
\DeclareRobustCommand{\vect}[1]{
  \ifcat#1\relax
    \boldsymbol{#1}
  \else
    \vb*{#1}
\fi}

\makeatletter
\newcommand{\linebreakand}{%
  \end{@IEEEauthorhalign}
  \hfill\mbox{}\par
  \mbox{}\hfill\begin{@IEEEauthorhalign}
}
\makeatother

\begin{document}
    \title{Malware families discovery via Open-Set Recognition on Android manifest permissions}
    
    \author{
        \IEEEauthorblockN{\hspace{-0.8cm}Filippo Leveni\IEEEauthorrefmark{1}\textsuperscript{(\Letter)}\orcidlink{0009-0007-7745-5686}\\ \tt\scriptsize \hspace{-0.8cm}filippo.leveni@polimi.it}
        \and
        \IEEEauthorblockN{Matteo Mistura\IEEEauthorrefmark{3}\\ \tt\scriptsize matteo.mistura@cleafy.com}
        \and
        \IEEEauthorblockN{Francesco Iubatti\phantom{\IEEEauthorrefmark{3}}\\ \tt\scriptsize \phantom{francesco.iubatti@cleafy.com}}
        \and
        \IEEEauthorblockN{Carmine Giangregorio\IEEEauthorrefmark{3}\\ \tt\scriptsize carmine.giangregorio@cleafy.com}
        \and
        \IEEEauthorblockN{Nicolò Pastore\IEEEauthorrefmark{3}\\ \tt\scriptsize nicolo.pastore@cleafy.com}
        \and
        \IEEEauthorblockN{Cesare Alippi\IEEEauthorrefmark{1}\IEEEauthorrefmark{2}\orcidlink{0000-0003-3819-0025}\\ \tt\scriptsize cesare.alippi@polimi.it\\cesare.alippi@usi.ch}
        \and
        \IEEEauthorblockN{Giacomo Boracchi\IEEEauthorrefmark{1}\orcidlink{0000-0002-1650-3054}\\ \tt\scriptsize giacomo.boracchi@polimi.it}
        \and
        \hspace{0.5cm}\IEEEauthorblockA{\IEEEauthorrefmark{1}\textit{Department of Electronics, Information and Bioengineering} \\
                          \textit{Politecnico di Milano}\\
                          Milan, Italy}
        \and
        \IEEEauthorblockA{\IEEEauthorrefmark{2}\textit{Faculty of Informatics} \\
                          \textit{Università della Svizzera Italiana}\\
                          Lugano, Switzerland}
        \and
        \IEEEauthorblockA{\IEEEauthorrefmark{3}\textit{\href{https://www.cleafy.com/}{Cleafy S.p.A.}} \\
                          Milan, Italy}
    }
    
    \maketitle
    \begin{abstract}
        Malware are malicious programs that are grouped into families based on their penetration technique, source code, and other characteristics. Classifying malware programs into their respective families is essential for building effective systems for defending against cyber threats. Machine learning models have a huge potential in malware detection on mobile devices, as malware families can be recognized by classifying permission data extracted from Android manifest files. Still, the malware classification task is challenging due to the high-dimensional nature of permission data and the limited availability of training samples. In particular, the steady emergence of new malware families makes it impossible to acquire a comprehensive training set covering all the malware classes. In this work, we present a malware classification system that, on top of classifying known malware, detects new ones. In particular, we combine an open-set recognition technique developed within the computer vision community, namely \maxlogit, with a tree-based \gboost classifier, which is particularly effective in classifying high-dimensional data. Our solution turns out to be very practical, as it can be seamlessly employed in a standard classification workflow, and efficient, as it adds minimal computational overhead. Experiments on public and proprietary datasets demonstrate the potential of our solution, which has been deployed in a business environment.
    \end{abstract}


    \section{Introduction}
        \label{sec:introduction}
        
        Malware programs are designed to disrupt, damage, or gain unauthorized access to computer systems or data. Android, the most used mobile device operating system~\cite{Statista24*1,Statista24*2}, has gained the attention of malware developers thanks to its popularity and increased use for business and financial activities.
        Malware programs have very different behaviors and characteristics, thus are classified into distinct families based on their attributes. Classifying malware is very important to enable quick identification and an effective response to potential attacks. Analyzing malware families also provides insights into evolving attack trends and patterns, informing security researchers with an updated view of the threat landscape. An accurate malware classification system is key to help organizations assess their risk exposure, prioritize security efforts, and effectively allocate resources to protect corporates and customers.
        
        \medskip
        Malware programs targeting Android devices often exploit permissions declared in the manifest file to gain unauthorized access to sensitive resources~\cite{OdusamiAbayomi-AlliAl18}. These malicious applications may request permissions beyond what is necessary for their stated functionality, tricking users into granting access to sensitive data or system resources. For example, malware programs can leverage permissions to carry out activities such as sending SMS messages, getting access to banking or payment app data, or downloading and executing additional malicious payloads. By abusing permissions declared in the manifest file, malware programs can operate stealthily, posing a significant threat to users' privacy and security.

        Malware classification often relies on rule-based approaches crafted by experts~\cite{OdusamiAbayomi-AlliAl18}, but these are ineffective in identifying complex patterns in the huge space of permission requests and are prone to introducing human biases. Conversely, machine learning models learn distinctive patterns directly from data, leading to a more accurate malware classification. Permission information is typically one-hot encoded, resulting in a very sparse and high-dimensional binary feature vector for each malware program. In such high-dimensional settings, tree-based classifiers like decision trees and gradient boosting machines are considered state-of-the-art, as they are particularly effective and efficient in capturing complex nonlinear relationships between binary features~\cite{HerronGlissonAl21,TurnipSitumorangAl20}.
        
        \medskip
        Malware developers continuously craft new malicious programs to elude classification by security systems. The emergence of new malware families poses significant challenges to cybersecurity, as these cannot be classified from a previously trained model. To ensure overall protection for Android devices, a robust malware classification system must effectively identify known classes and also detect unknown ones.
        The latter, known as Open-Set Recognition (\osr), is typically addressed by assessing the classifier's uncertainty. \osr has been widely addressed within the computer vision community, thus most solutions are tailored for neural network classifiers~\cite{BendaleBoult16,GeDemyanov17,LeeLeeAl18}, which are not well suited for malware classification. To the best of our knowledge, \osr methods have seldom been extended to tree-based classifier~\cite{FengDesaiAl24}, that are widely used in malware classification.
        
        \medskip
        In this work, we present a practical and efficient solution for identifying unknown malware families, namely those not represented in the training set. Specifically, we extend \maxlogit, a simple yet widely used \osr technique developed for neural networks, to tree-based \gboost classification models. In contrast with recent malware \osr solutions, which involve sophisticated deep learning architectures often requiring ad-hoc training procedures, such as Generative Adversarial Networks (GANs)~\cite{GuoGuoAl23}, Transformer-based models~\cite{LuWang24}, or multimodal deep embeddings~\cite{GuoWangAl25}, our solution can be seamlessly integrated into an existing tree-based malware classifier, without even modifying the training procedure. The above-cited methods poorly fit with real-world industrial scenarios as they often require large volumes of training data, high computational costs, complex training procedures, and lack of compatibility with existing systems.
        
        Specifically, we show that \maxlogit can be readily applied on decision values already produced by a \gboost classifier, enabling \osr on tree-based models -- widely adopted in malware detection systems -- without requiring any changes to existing pipelines. Our solution is, therefore, tailored for resource-constrained industrial environments, prioritizing low latency, data efficiency, and ease of integration with existing infrastructures. We validate our solution on both public and proprietary real-world datasets, demonstrating its effectiveness and superior performance over a nearest-neighbor \osr baseline~\cite{MendesDeSouzaAl17} in handling high-dimensional binary data, which can also be seamlessly integrated into a pre-trained tree-based classifier, thus is a viable alternative to ours. We further investigate the nature of false alarms generated by the open-set recognition module, showing that a substantial fraction originates from samples that are already ambiguous or incorrectly classified by the underlying closed-set model. This finding suggests that the proposed \osr module acts not only as a detector of previously unseen malware families, but also as a mechanism for flagging uncertain and potentially high-risk samples for manual review, reducing the practical impact of incorrect classifications in operational deployments. Importantly, our solution has been successfully deployed in a business environment, and it is currently part of their malware classification engine. Preliminary evidence gathered through its day-to-day use in threat intelligence monitoring confirms the practical utility of the proposed approach and is consistent with the experimental results reported in this paper.

    \section{Background}
        \label{sec:background}

        Malware classification approaches typically rely on heuristic rules, and can be divided in static, such as signature-based and permission-based, and dynamic~\cite{OdusamiAbayomi-AlliAl18}.
        Dynamic analysis involves observing application behavior within a sandbox environment, and monitoring system calls to construct a function call graph which is analyzed to identify malicious behaviour.
        Some dynamic approaches employ machine learning algorithms to classify malware programs from their function call graphs~\cite{HassenChan17}. Depending on the employed classifier, it can be easy to integrate an open-set recognition technique to discover new malware families~\cite{HassenChan20,JiaChan22*1}. Despite the high potential of dynamic analysis, the runtime testing required to construct the function call graph makes it inefficient and unsuitable for high throughput scenarios.
        
        In signature-based approaches, unique identifiers of known malware programs are stored in a database, and any application is compared to them and eventually flagged as malware~\cite{FarukiLaxmiAl15}. Signature-based analysis proves efficient, but it is effective only for malware families that are already stored in the database, making it not appropriate to identify new ones.
        
        Permission analysis, on the other hand, consists in classifying programs based solely on the permissions they request from the operating system~\cite{RovelliVigfusson14}. In this work, we focus on permission analysis as it is more flexible than signature-based analysis and more efficient compared to dynamic analysis.

        \medskip
        Android applications specify required permissions in a mandatory file named AndroidManifest, which includes both custom and system permissions. Custom permissions do not require access to sensitive data such as contacts or filesystem, whereas system permissions encompass all permissions exposed by the system, with only the most sensitive ones requiring explicit user approval. Our study focuses solely on system permissions due to their potential security risks for the user.
        
        After filtering out custom permissions and applying one-hot encoding to system permissions, it is possible to detect and classify malware programs through a machine learning model. After one-hot encoding, each permission is represented by a binary value ($1$ for requested, $0$ for not requested), thus each application is represented by a high-dimensional binary vector of length $P$, where $P$ is the total number of system permissions. However, the limited number of samples typically available in real-world malware classification scenarios (small $n$), coupled with the high-dimensional feature space (large $P$) poses challenges for effectively training classification models.

    \section{Problem formulation}
        \label{sec:problem_formulation}
        
        A malicious application is represented by a vector of permissions $\vect{p} \in \{0, 1\}^P$, where $P$ is the total number of permissions considered, and $p_i = 1$ when the $i$-th permission is listed in the application manifest.
        These malicious applications might either belong to a known class $\ell \in \mathcal{L}$ indicating a known malware family, or to a novel family that has never been observed before.

        Our goal is to train an open-set classifier $\mathcal{K}$ that associates to each malicious application $\vect{p}$ either a known class label $\widehat{\ell}(\vect{p}) \in \mathcal{L}$ or the \emph{Novel} label, \emph{i.e.}:
        \begin{equation}
            \label{eq:osr_classifier}
            \mathcal{K}(\vect{p}) =
            \begin{cases}
                \textit{Novel}\\
                \widehat{\ell}(\vect{p}) \in \mathcal{L}.
            \end{cases}
        \end{equation}
        We assume that we are provided with a training set $\mathcal{TR} = \{(\vect{p}_i, \ell_i) \;|\; \ell_i \in \mathcal{L}\}_{i = 1, \dots, n}$ of annotated malicious applications belonging to known families and with a test set $\mathcal{TS} = \{(\vect{p}_i, \ell_i) \;|\; \ell_i \in \mathcal{L} \cup \{Novel\}\}_{i = 1, \dots, m}$ of annotated malicious applications belonging to both known and novel families.

    \section{Related work}
        \label{sec:related_work}
        
        Using the manifest file permissions alongside machine learning models for malware classification is a well-established practice. In~\cite{RovelliVigfusson14,HerronGlissonAl21}, the permissions vector serves as behavioral marker and it is fed to machine learning models ranging from Support Vector Machines and Gaussian Naive Bayes to Random Forests. Their findings indicate that machine learning models trained solely on manifest file permissions significantly outperform traditional anti-virus engines, with Random Forest achieving the highest accuracy. In~\cite{TurnipSitumorangAl20}, a tree-based \gboost model classifies six malware classes using three permission categories, highlighting that boosting is particularly suited for malware classification.
        However, all these methods operate as closed-set classifiers and are therefore unable to discover new malware families.
        
        \medskip
        An effective machine learning malware classification system should both accurately classify known malware programs and identify novel, unknown malware families. This challenge, known as \osr~\cite{BendaleBoult16}, is typically addressed by assessing the classifier's uncertainty.
        \osr systems can be categorized into two types. The first type distinguishes between instances of known families and unknown ones, but does not differentiate among known families~\cite{ScheirerRochaAl12,BodesheimFreytagAl13,BodesheimFreytag15}. This kind of \osr approach, also referred to as \emph{anomaly detection}, does not classify known malware programs, which is a primary requirement in our setting. Conversely, the second type of \osr systems can both classify known families and identify instances of unknown ones~\cite{BendaleBoult16,DaYuAl14}.
        
        The \osr problem has been extensively studied in the computer vision community~\cite{BendaleBoult16,GeDemyanov17,LeeLeeAl18,VazeHanAl21}, with most solutions designed for convolutional neural networks.
        Neural networks and \osr have been used in dynamic analysis for malware detection~\cite{HassenChan20,JiaChan22*1}, where applications are executed in a sandbox to extract function call graphs, converted into adjacency matrices. A convolutional neural network then classifies these matrices identifying discriminative patterns of malware families in a latent space. During testing, an instance is classified as unknown when the distance to its closest centroid exceeds a threshold.
        While effective, the extraction of function call graphs is time-consuming, making it impractical for high throughput scenarios.
        Static analysis for malware detection have also been extended to \osr.
        In~\cite{ZhouWang12} a permission-based footprinting is used to classify samples of known Android malware families and then a heuristics-based filtering identifies specific behaviors exhibited by unknown malicious families. When an application classified as malicious does not appear in the database, they consider it as novel and generate the corresponding permission-based footprint in a feedback loop. However, this heuristic-based approach targets only specific Android features that may be exploited to load new code, thus limiting the ability to identify different types of unknown malware programs.
        
        Recent advances in \osr for malware detection have moved toward increasingly complex deep learning architectures. For instance, in~\cite{GuoGuoAl23} a Conservative Novelty Synthesizing Network based on GANs generates malware samples that help distinguish unknown families. DOMR~\cite{LuWang24} is a Transformer-based \osr method that relies on episodic training and meta-learning, while in~\cite{GuoWangAl25} \osr is extended to multimodal settings using dual-embedding networks combining CNNs and BERT-like Transformers. Although these methods report strong performance on large-scale datasets, they rely heavily on abundant labeled data, high-end computing resources, and ad-hoc training procedures, making them unsuitable and unpractical for many industrial environments.
        
        In industrial environments, it is preferable to address \osr using models trained through efficient procedures that require minimal changes to existing closed-set classifiers, facilitating their deployment and maintenance. A lightweight example is Open-Set Nearest-Neighbor (\osnn)~\cite{MendesDeSouzaAl17}, which extends a $1$-NN classifier to address the \osr task. \osnn computes the ratio of distances between a sample and its two nearest neighbors from different families and classifies as unknown all the samples having ratio below a specified threshold. However, our experiments show that \osnn performs poorly on high-dimensional data.
        Tree-based models like \gboost, on the other hand, are widely used in cybersecurity pipelines due to their efficiency and strong performance on high-dimensional sparse data -- like one-hot encoded Android permissions -- but have seldom been extended to \osr~\cite{FengDesaiAl24}.
        
        \medskip
        In this work, we combine the effectiveness and efficiency of \gboost~\cite{HastieTibshiraniAl09} in classifying malware programs, with the \maxlogit \osr technique, originally developed within the computer vision community~\cite{VazeHanAl21}. To the best of our knowledge, \osr techniques have seldom been applied to tree-based classifiers, and by extending \maxlogit to operate with \gboost, our work contributes in filling this gap by providing an \osr solution that is practically feasible in real-world settings.

    \section{Proposed approach}
        \label{sec:proposed_approach}
        
        \begin{figure}[t]
            \centering
            \includegraphics[width=\linewidth]{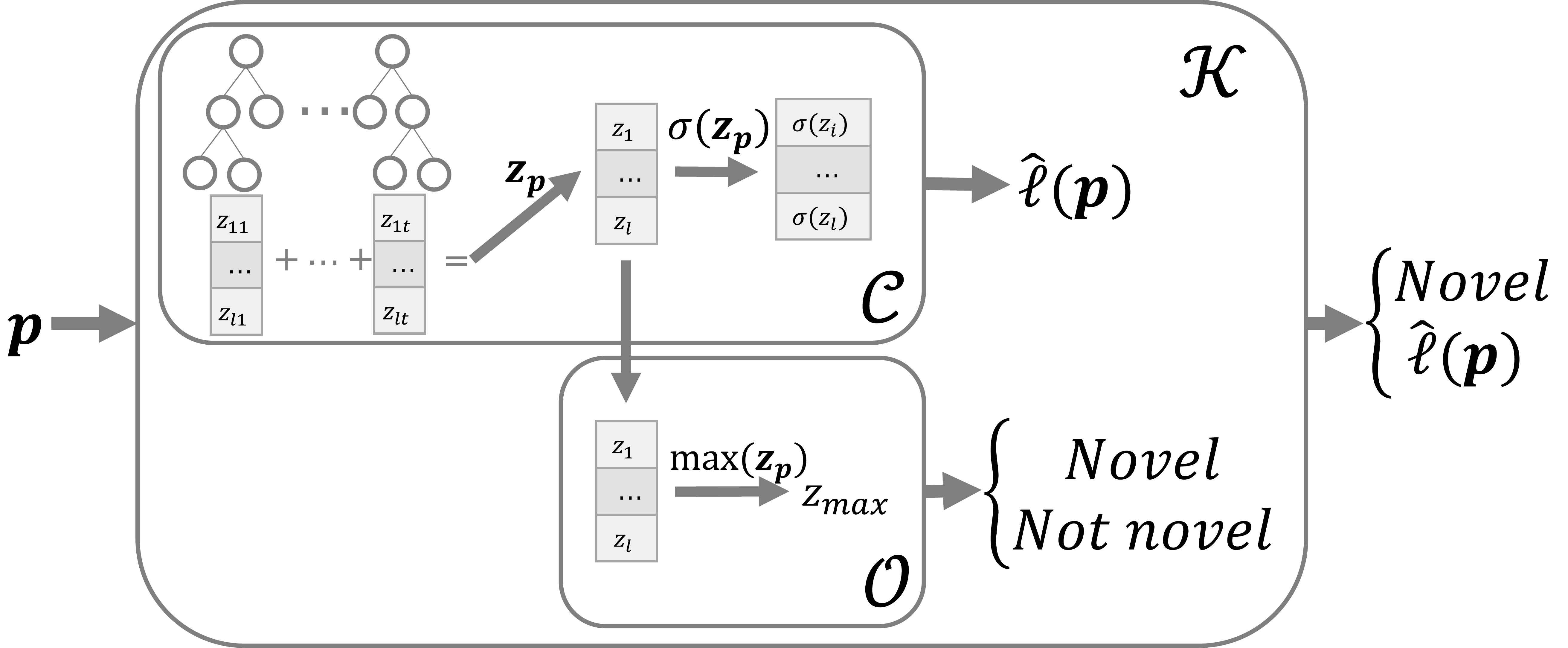}
            \vspace{-0.2cm}
            \caption{\footnotesize Depiction of our proposed open-set classifier $\mathcal{K}$.}
            \vspace{0.8cm}
            \label{fig:open_set_classifier}
        \end{figure}

        Our proposed open-set classifier $\mathcal{K}$, depicted in~\cref{fig:open_set_classifier}, consists of a closed-set classifier $\mathcal{C}$ and an open-set recognition module $\mathcal{O}$. The closed-set classifier assigns to each input malicious application $\vect{p}$ a known class label, expressed as $\mathcal{C}(\vect{p}) = \widehat{\ell}(\vect{p}) \in \mathcal{L}$. The open-set recognition module $\mathcal{O}(\vect{z}_{\vect{p}}) \in \{Novel, Not\;novel\}$ determines whether $\vect{p}$ belongs to a $novel$ (unknown) or $not$ $novel$ (known) class based on the raw output values $\vect{z}_{\vect{p}}$ assigned to $\vect{p}$ by the classifier $\mathcal{C}$. Therefore, we can address problem~\eqref{eq:osr_classifier} in the following way:
        \begin{equation*}
            \mathcal{K}(\vect{p}) =
            \begin{cases}
                \textit{Novel} \quad \textnormal{if} \; \mathcal{O}(\vect{z}_{\vect{p}}) = Novel\\
                \mathcal{C}(\vect{p}) \quad\;\;\textnormal{otherwise}.
            \end{cases}
        \end{equation*}
        We employ tree-based \gboost~\cite{HastieTibshiraniAl09} as closed-set classifier, that is a set of boosted classification trees $\mathcal{C} = \{T_i\}_{i = 1, \dots, t}$. The \maxlogit~\cite{VazeHanAl21} as open-set recognition module $\mathcal{O}$ takes as input the logit vector $\vect{z}_{\vect{p}}$ of the classifier $\mathcal{C}$ and leverages a threshold $\tau \in \mathbb{R}$ that ensures a specified false alarm rate.
        
        \subsection{\logit extraction}
            \label{subsec:logit_extraction}

            \begin{figure}[t]
                \centering
                \begin{subfigure}[t]{\linewidth}
                    \centering
                    \includegraphics[width=\linewidth]{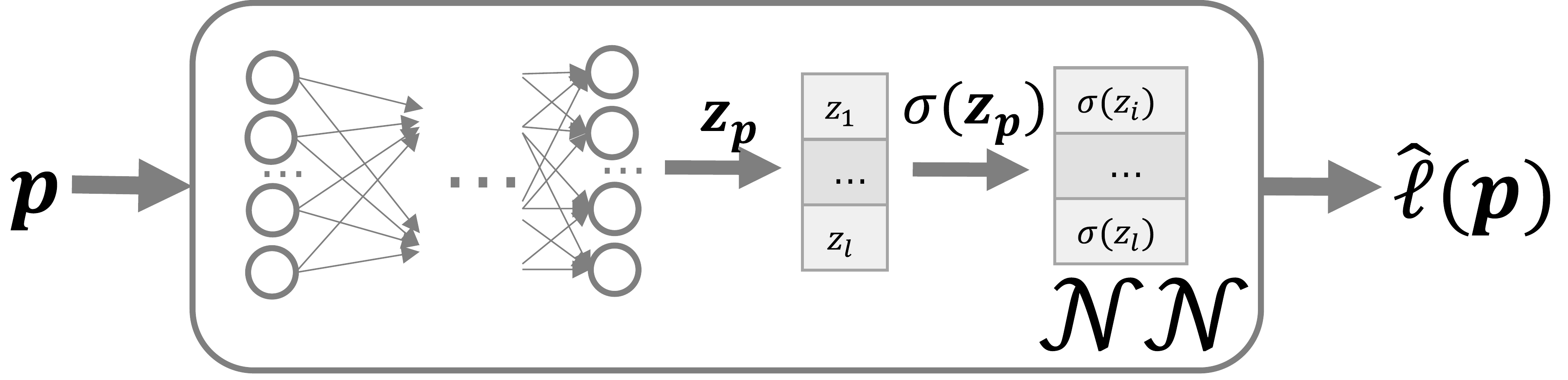}
                    \caption{}
                    \label{fig:nn_maxlogit}
                \end{subfigure}
                \begin{subfigure}[t]{\linewidth}
                    \vspace{.25cm}
                    \centering
                    \includegraphics[width=\linewidth]{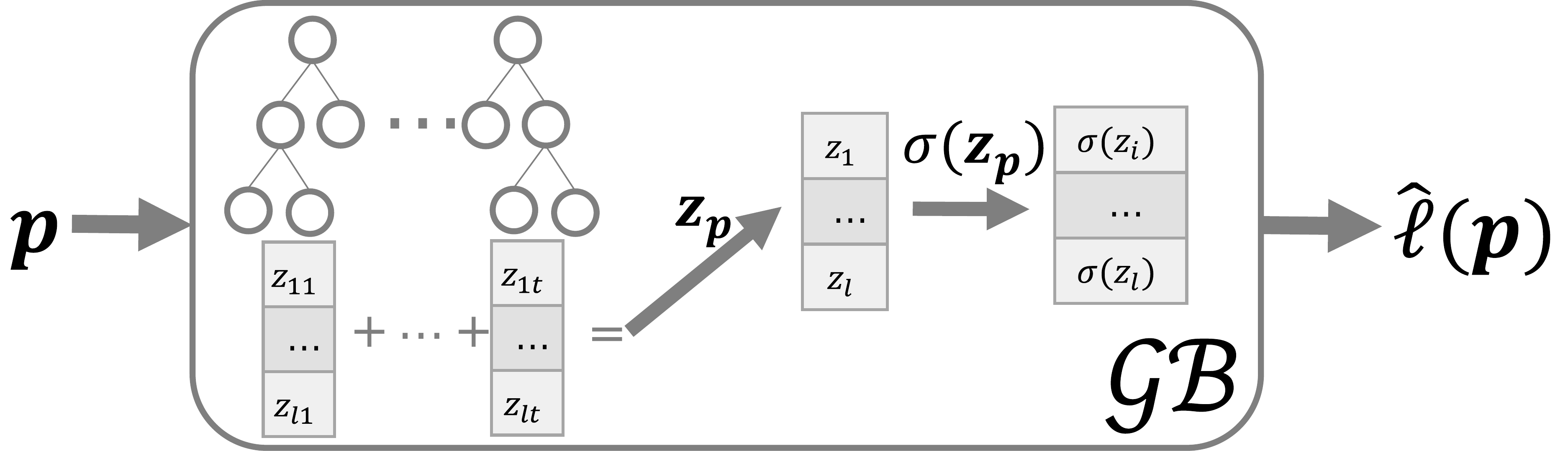}
                    \caption{}
                    \label{fig:gb_maxlogit}
                \end{subfigure}
                \vspace{0.3cm}
                \caption{\footnotesize Analogy between the \logits vector in a neural network-based classifier (a) and the decision values in a tree-based \gboost classifier (b). In both cases, the vector $\vect{z}_{\vect{p}}$ represents the raw output scores before applying the softmax function $\sigma(\vect{z}_{\vect{p}})$.}
                \vspace{0.8cm}
                \label{fig:maxlogit}
            \end{figure}
            
            \maxlogit has been originally developed within the computer vision community and has traditionally been coupled with neural network-based classifiers. In this context, \maxlogit operates on the raw output scores $\vect{z}_{\vect{p}} = [z_1, \dots, z_l]$ generated by the last layer of the neural network, known as the \logits, before the softmax function. In~\cref{fig:nn_maxlogit} we illustrated a neural network classifier, where $\vect{p}$ is the input point and $\widehat{\ell}(\vect{p})$ is the predicted label.
            Since the softmax does not change the order of the scores in $\vect{z}_{\vect{p}}$, the neuron containing the maximum logit score corresponds to the predicted class $\widehat{\ell}(\vect{p})$. The softmax function $\sigma(\vect{z}_{\vect{p}}) = [\sigma(z_1), \dots, \sigma(z_l)]$ operates on the \logits vector $\vect{z}_{\vect{p}}$ as follows:
            \begin{equation}
                \label{eq:softmax}
                \sigma(z_i) = \frac{e^{z_i}}{\sum_{j = 1}^{l}e^{z_j}}.
            \end{equation}
            \maxlogit approach operates directly on the raw output values $\vect{z}_{\vect{p}}$ because \logits display better separation between known and unknown classes when compared to normalized probability values $\sigma(\vect{z}_{\vect{p}})$~\cite{VazeHanAl21}.
            
            Unfortunately, neural networks are not well suited for handling high-dimensional sparse binary data. Hence, we opted for tree-based \gboost as the closed set classifier $\mathcal{C}$ due to its effectiveness in managing such data. To the best of our knowledge, this is the first time that the \maxlogit approach is used in conjunction with classifiers other than those based on neural networks.
            A \gboost classifier does not have proper \logits, but it produces unnormalized scores assigned to each class by the ensemble model before being transformed into probabilities (\cref{fig:gb_maxlogit}).
            In \gboost, in order to obtain the final prediction $\widehat{\ell}(\vect{p})$ for a given input sample $\vect{p}$, decision values are aggregated by summation across all the trees in the ensemble, resulting in a final vector $\vect{z}_{\vect{p}} = [z_1, \dots, z_l]$. Each $z_i = \sum_{j = 1}^{t} z_{ij}$ is obtained by summing the decision values for the $i$-th label along all the trees, as illustrated in~\cref{fig:gb_maxlogit}. Subsequently, similarly to neural network classifiers, the softmax~\eqref{eq:softmax} is applied to $\vect{z}_{\vect{p}}$, and the class corresponding to the highest probability in $\sigma(\vect{z}_{\vect{p}})$ is designated as the predicted class for the input sample $\vect{p}$.
            Our intuition is to use \gboost decision values as the \logits vector produced by neural networks.
            
        \subsection{\maxlogit computation}
            \label{subsec:maxlogit_computation}
            
            The \maxlogit open-set recognition module $\mathcal{O}$ operates as a binary classifier, employing $\vect{z}_{\vect{p}}$, the maximum \logit value obtained from $\mathcal{C}$, to distinguish between $novel$ and $known$ classes.
            When a sample $\vect{p}$ is put into the \gboost classifier $\mathcal{C}$, we extract the \logits vector $\vect{z}_{\vect{p}} = [z_1, \dots, z_l]$ and identify the maximum value $\max(\vect{z}_{\vect{p}}) = \max(z_1, \dots, z_l)$. To decide whether a sample belongs to a novel class, we apply a threshold $\tau$ to $\max(\vect{z}_{\vect{p}})$ and, if $\max(\vect{z}_{\vect{p}}) < \tau$, it indicates that the classifier is not confident in its classification, thereby we classify $\vect{p}$ as $novel$:
            \begin{equation*}
                \mathcal{O}(\vect{z}_{\vect{p}}) =
                \begin{cases}
                    \textit{Novel} \qquad\quad \textnormal{if} \; \max(\vect{z}_{\vect{p}}) < \tau\\
                    \textit{Known} \quad\;\textnormal{otherwise}.
                \end{cases}
            \end{equation*}

            The value of the threshold $\tau$ is fundamental to control false alarms raised by the \osr module, \emph{i.e.}, the amount of samples belonging to $known$ malware families classified as $novel$. Since the classification of a sample $\vect{p}$ as belonging to a new family leads to a subsequent manual inspection by a human expert, tuning the threshold $\tau$ is essential to prevent wasting human resources.
            Threshold $\tau$ is typically tuned using an external training set $\mathcal{TT} = \{(\vect{p}_i, \ell_i) \;|\; \ell_i \in \mathcal{L}\}_{i = 1, \dots, k}$, which is composed only of samples belonging to known classes. This is due to the fact that the false positive rate is solely impacted by misclassifications of known samples as novel.
            In particular, given a trained closed-set classifier $\mathcal{C}$, an external training set $\mathcal{TT}$ and a target false positive rate $FPR \in [0, 1]$, we set the threshold $\tau$ such that:
            \begin{equation*}
                \mathcal{P}(\max(\vect{z}_{\vect{p}}) < \tau \;|\; \vect{p} \in \mathcal{TT}) \leq FPR
            \end{equation*}
            where $\mathcal{P}$ denotes probability. In practice, it is enough to set the threshold $\tau$ equal to the $FPR$-quantile of the empirical distribution of $\max(\vect{z}_{\vect{p}})$ computed from the elements of $\mathcal{TT}$.

            \medskip
            We remark that our method does not add any complexity to the \gboost closed-set classifier $\mathcal{C}$. The open-set recognition module $\mathcal{O}$ is seamlessly integrated into the classification workflow, preserving an inference time complexity of $O(t \: d)$ per sample, where $t$ is the number of trees in the ensemble and $d$ is the maximum tree depth.

    \section{Experiments}
        \label{sec:experiments}
        
        In this section, we assess the benefits of our open-set recognition solution in identifying new malware families on both a publicly available dataset and a proprietary dataset provided by \cleafy. We first introduce the datasets in~\cref{subsec:dataset}, and detail the experimental methodology in~\cref{subsec:methodology}. Subsequently, we validate our approach for the open-set recognition task on the considered datasets, compare it against a representative \osr baseline, and analyze the nature of the false alarms generated by the proposed method. In particular, we investigate whether samples flagged as novel correspond to genuinely uncertain malware instances, thereby providing additional value to threat analysts beyond the discovery of previously unseen families. Finally, we discuss the performance of our solution within \cleafy's real-world deployment settings in~\cref{subsec:results_and_discussion}.
        
        \subsection{Dataset}
            \label{subsec:dataset}
            
            \begin{figure}[t]
                \centering
                \includegraphics[width=.85\linewidth]{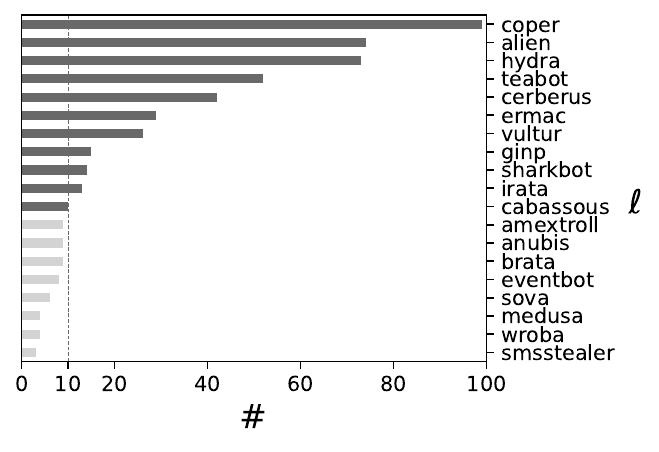}
                \vspace{-0.1cm}
                \caption{\footnotesize Imbalance in the cardinality $\#$ of each malware family $\ell$ in our proprietary dataset, where malware families with less than $10$ samples are represented by pale-colored bars.}
                \vspace{0.8cm}
                \label{fig:families_cardinality}
            \end{figure}
            
            We use the public dataset Drebin~\cite{ArpSpreitzenbarthAl14}, which consists of $n = 5560$ applications from $179$ different malware families, collected in the period from August 2010 to October 2012. Each application is characterized by several features grouped into the following categories: \emph{hardware components}, \emph{required permission}, \emph{app components}, \emph{filtered intents}, \emph{restricted API calls}, \emph{used permission}, \emph{suspicious API calls} and \emph{network addresses}. Since this study focuses on system permissions, we considered only the \emph{required permission} group explicitly referring to the Android operating system.
            
            The proprietary dataset, kindly provided by \cleafy, comprises $n = 499$ system permission vectors extracted from the manifest of malicious applications, identified as a subset of the threat intelligence telemetry provided by an online cybersecurity software. The considered telemetry contains applications identified as potentially malicious on user devices worldwide, with a particular focus on Europe, in the period from February 2022 to January 2023. The malware labels were assigned through manual analysis of malware programs conducted by a highly specialized threat intelligence team with strong domain expertise. \cref{fig:families_cardinality} illustrates the various malware families present in the proprietary dataset and their cardinalities, highlighting the imbalanced nature of the classification problem.
            
            Each application is described by a binary vector $\vect{p}_i \in \{0, 1\}^P$, where $P$ is $154$ and $1800$ for the public and proprietary dataset respectively, containing the one-hot encoding of the requested Android permissions, along with the corresponding label $\ell_i \in \mathcal{L}$ denoting the malware family. The resulting datasets are in the form of $\mathcal{D} = \{(\vect{p}_i, \ell_i) \;|\; \ell_i \in \mathcal{L}\}_{i = 1, \dots, n}$.

        \subsection{Methodology}
            \label{subsec:methodology}

            We evaluate the closed-set recognition performance of \gboost, on both public and proprietary datasets, through stratified $10$-fold cross-validation and average results over the $10$ test folds. Malware families with less than $10$ samples were excluded from the $10$-fold validation procedure and grouped in a dummy class labeled as $others$, which we then subsequently employed as the $novel$ class for assessing the open-set recognition performance. This results in an open-set recognition problem with $54$ and $11$ known malware families for Drebin and proprietary dataset respectively, and in $125$ and $8$ novel families respectively.
            On the public dataset, we also grouped all the malware families other than the top $10$ most populous ones into the $others$ class, and we refer to this configuration as Drebin\textsubscript{10}. This setup (\fixnovel) allows us to investigate a scenario where the closed-set classifier $\mathcal{C}$ has to deal with a low number of frequent families, while the open-set recognition module $\mathcal{O}$ has to identify a larger, more heterogeneous set of samples as $novel$.
            
            We perform an additional experiment following a leave-one-class-out approach (\loonovel), by training the closed set classifier $\mathcal{C}$ on $9$ out of the $10$ frequent classes of Drebin\textsubscript{10}. The excluded malware class is considered $novel$ at test time. By doing so, we evaluate the effectiveness of our model in recognizing each malware class as novel when trained on the other classes. We do not consider leave-$k$-class-out procedures with $k > 1$, as increasing $k$ reduces the number of known classes, thereby simplifying both the closed-set and open-set recognition tasks. The performance gap between Drebin and Drebin\textsubscript{10} supports this design choice. We employ stratified $10$-fold cross-validation within each leave-one-class-out iteration, and average the results over the $10$ test folds. To further assess the capability of our solution in identifying each individual malware class as $novel$, we plot the ROC curves for each novel class detection problem, treating the novel malware class as the positive class and merging all the classes used for training into the negative class.
            
            \medskip
            Unfortunately, we have no external training set to estimate the threshold $\tau$, nor can we use a portion of our datasets $\mathcal{D}$ solely for this purpose due to their limited size. Therefore, we set a desired false positive rate $FPR = 0.005$ and tuned the threshold $\tau$ on \emph{training} data, aware that this results in an underestimate of the real false positive rate at test time. A very low $FPR$ is mandatory in our settings, as classifying a sample $\vect{p}$ as belonging to a new family triggers manual inspection, which incurs a significant human resource cost.

        \subsection{Results and discussion}
            \label{subsec:results_and_discussion}
            
            \begin{table}[t]
                \caption{\footnotesize Micro-average (accuracy) and macro-average recall for closed-set $\mathcal{C}$ and open-set $\mathcal{K}$ classifiers on both public and proprietary datasets. The results are shown for the two tested settings \fixnovel (a) and \loonovel (b).}
                \vspace{0.4cm}
                \begin{subtable}{.49\textwidth}
                    \footnotesize
                    \centering
                    \caption{\footnotesize Less popolous malware families grouped into a dummy class $others$ and considered as the $novel$ class.}
                    \begin{tabular}{l|c|c|c|c}
                                                      & \multicolumn{2}{c|}{$\mathcal{C}$} & \multicolumn{2}{c}{$\mathcal{K}$} \\ \cline{2-5}
                                                      & Micro   & Macro                    & Micro   & Macro                   \\ \hline
                        Drebin\textsubscript{10}      & $0.931$ & $0.949$                  & $0.711$ & $0.874$                 \\
                        Drebin                        & $0.823$ & $0.856$                  & $0.772$ & $0.839$                 \\
                        Proprietary                   & $0.863$ & $0.869$                  & $0.842$ & $0.844$                 \\
                    \end{tabular}
                    \label{tab:kfold}
                \end{subtable}
                \\\\
                \begin{subtable}{.49\textwidth}
                    \footnotesize
                    \centering
                    \vspace{0.1cm}
                    \caption{\footnotesize Each populous malware family, in turn, designated as the $novel$ class in a leave-one-class-out fashion.}
                    \begin{tabular}{l|c|c|c|c}
                                                      & \multicolumn{2}{c|}{$\mathcal{C}$} & \multicolumn{2}{c}{$\mathcal{K}$} \\ \cline{2-5}
                                                      & Micro   & Macro                    & Micro   & Macro                   \\ \hline
                        Drebin\textsubscript{10}      & $0.935$ & $0.953$                  & $0.858$ & $0.874$                 \\
                        Drebin                        & $0.825$ & $0.850$                  & $0.810$ & $0.835$                 \\
                        Proprietary                   & $0.871$ & $0.881$                  & $0.858$ & $0.860$                 \\
                    \end{tabular}
                    \label{tab:loco_kfold}
                \end{subtable}
                \vspace{0.2cm}
                \label{tab:kfold_loco_kfold}
            \end{table}
            
            \begin{figure*}[t]
                \centering
                \begin{subfigure}[t]{.33\linewidth}
                    \centering
                    \includegraphics[width=\linewidth]{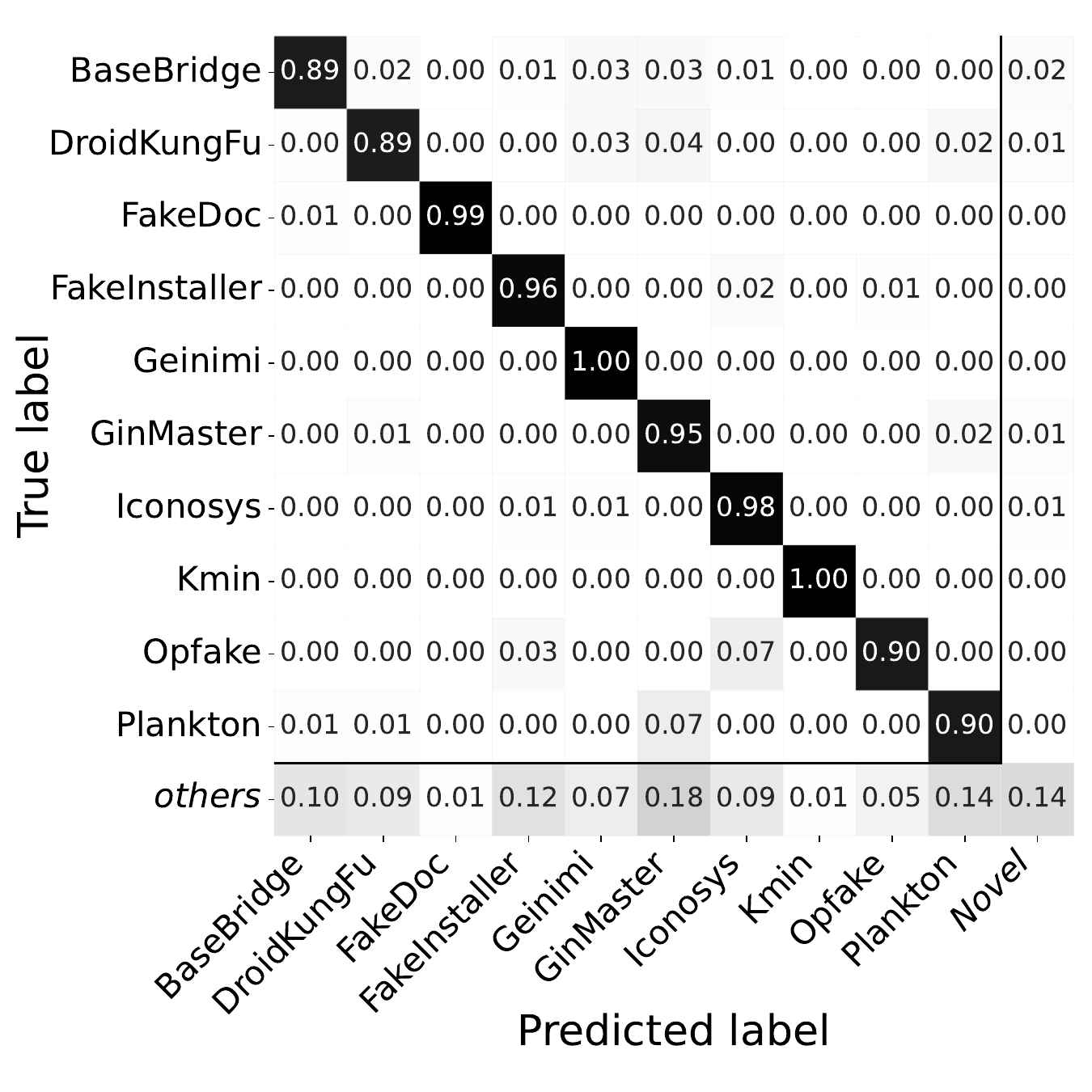}
                    \caption{}
                    \label{fig:kfold_osr_recall_matrix_public}
                \end{subfigure}
                \hfill
                \begin{subfigure}[t]{.33\linewidth}
                    \centering
                    \includegraphics[width=\linewidth]{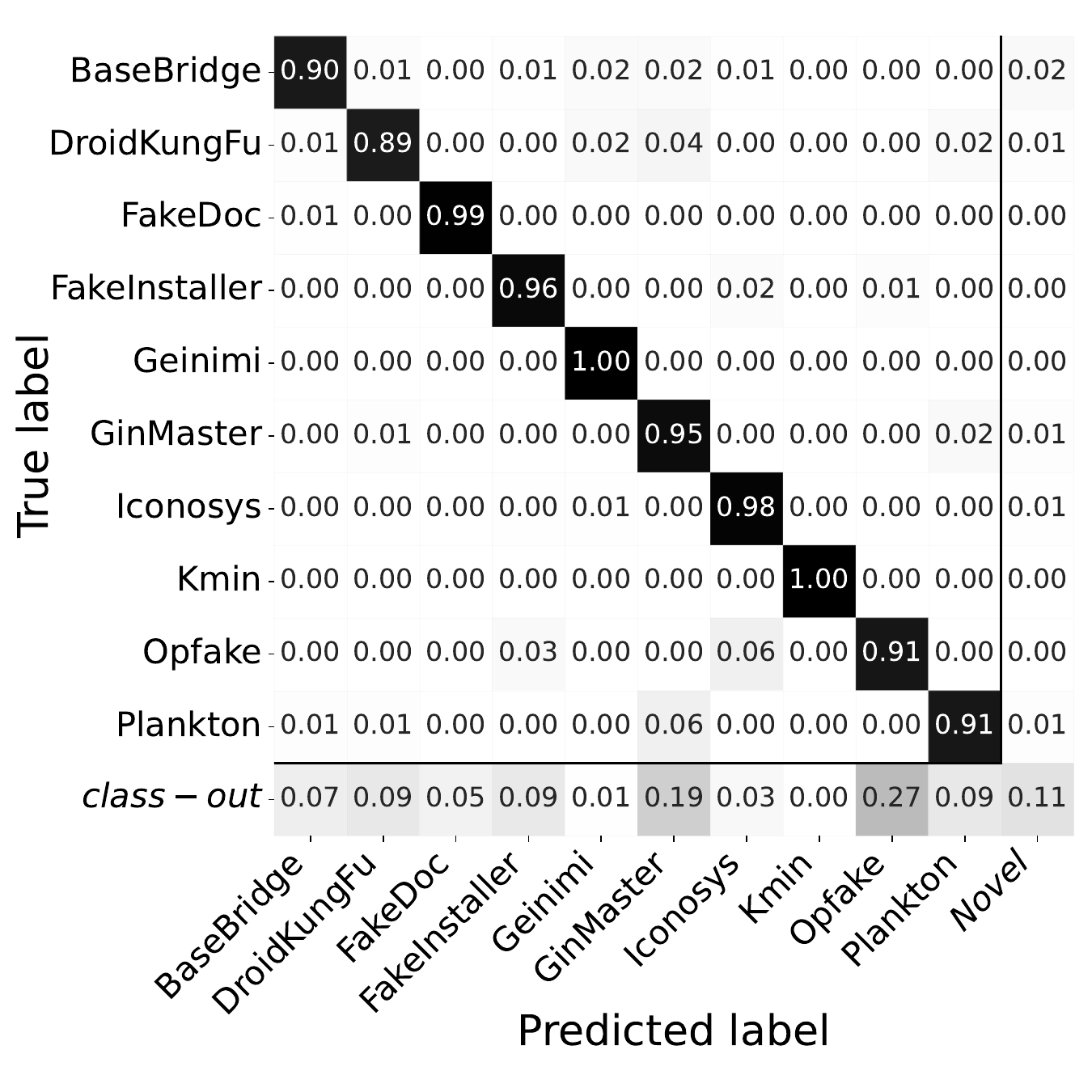}
                    \caption{}
                    \label{fig:loco_kfold_osr_recall_matrix_public}
                \end{subfigure}
                \hfill
                \vspace{0.3cm}
                \begin{subfigure}[t]{.28\linewidth}
                    \centering
                    \raisebox{7.7mm}{\includegraphics[width=\linewidth]{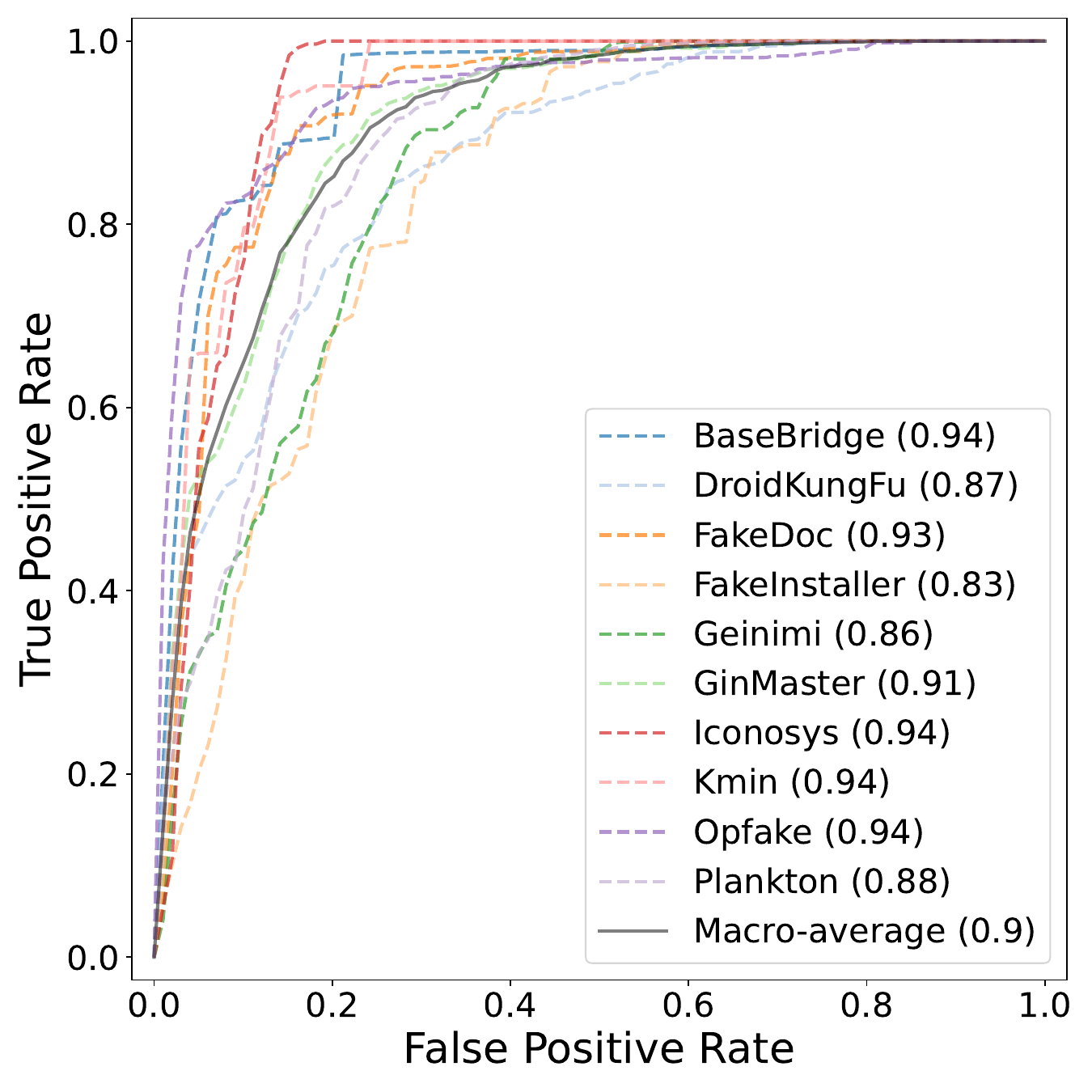}}
                    \caption{}
                    \label{fig:loco_kfold_osr_binary_roc_curves_public}
                \end{subfigure}
                
                \begin{subfigure}[t]{.33\linewidth}
                    \centering
                    \includegraphics[width=\linewidth]{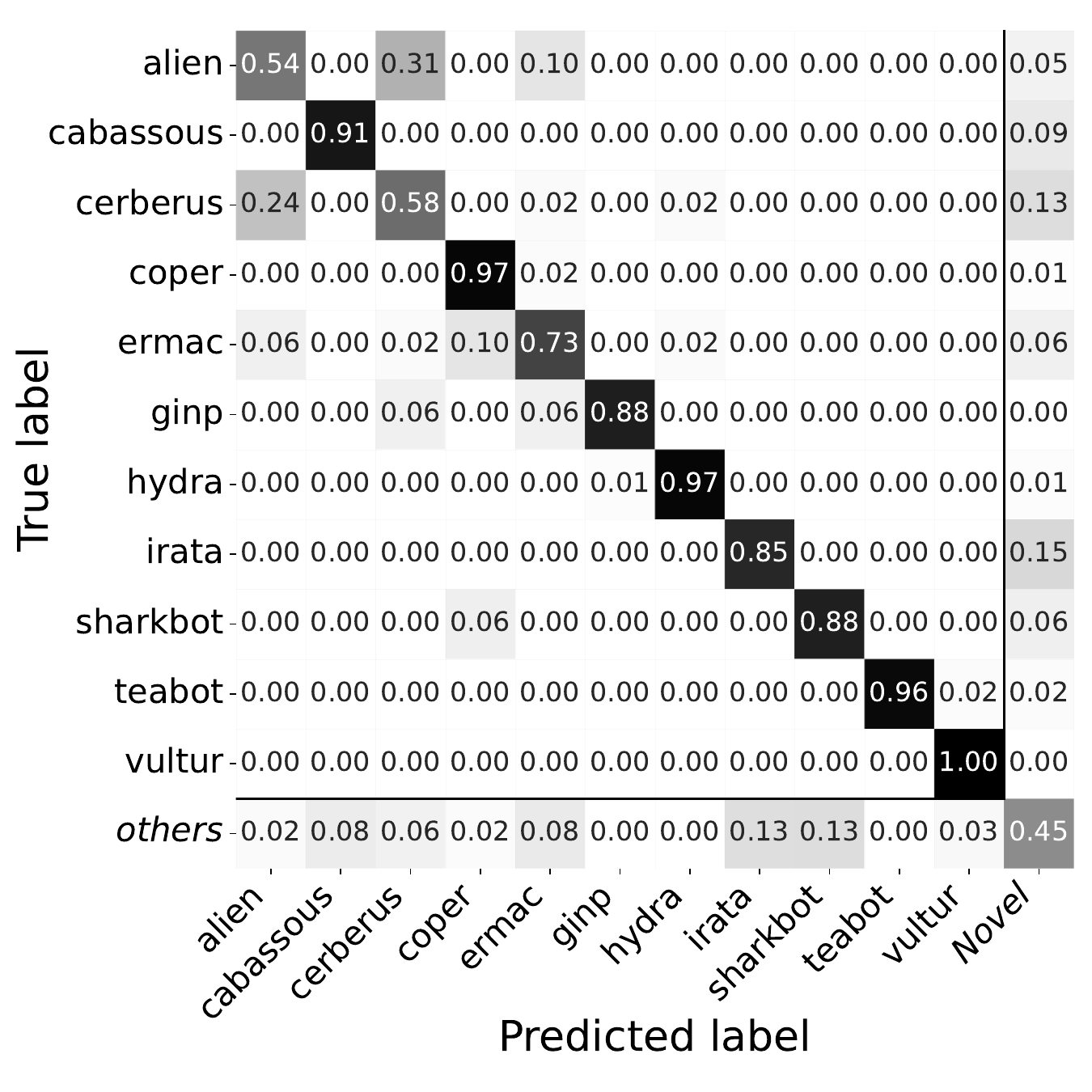}
                    \caption{}
                    \label{fig:kfold_osr_recall_matrix_private}
                \end{subfigure}
                \hfill
                \begin{subfigure}[t]{.33\linewidth}
                    \centering
                    \includegraphics[width=\linewidth]{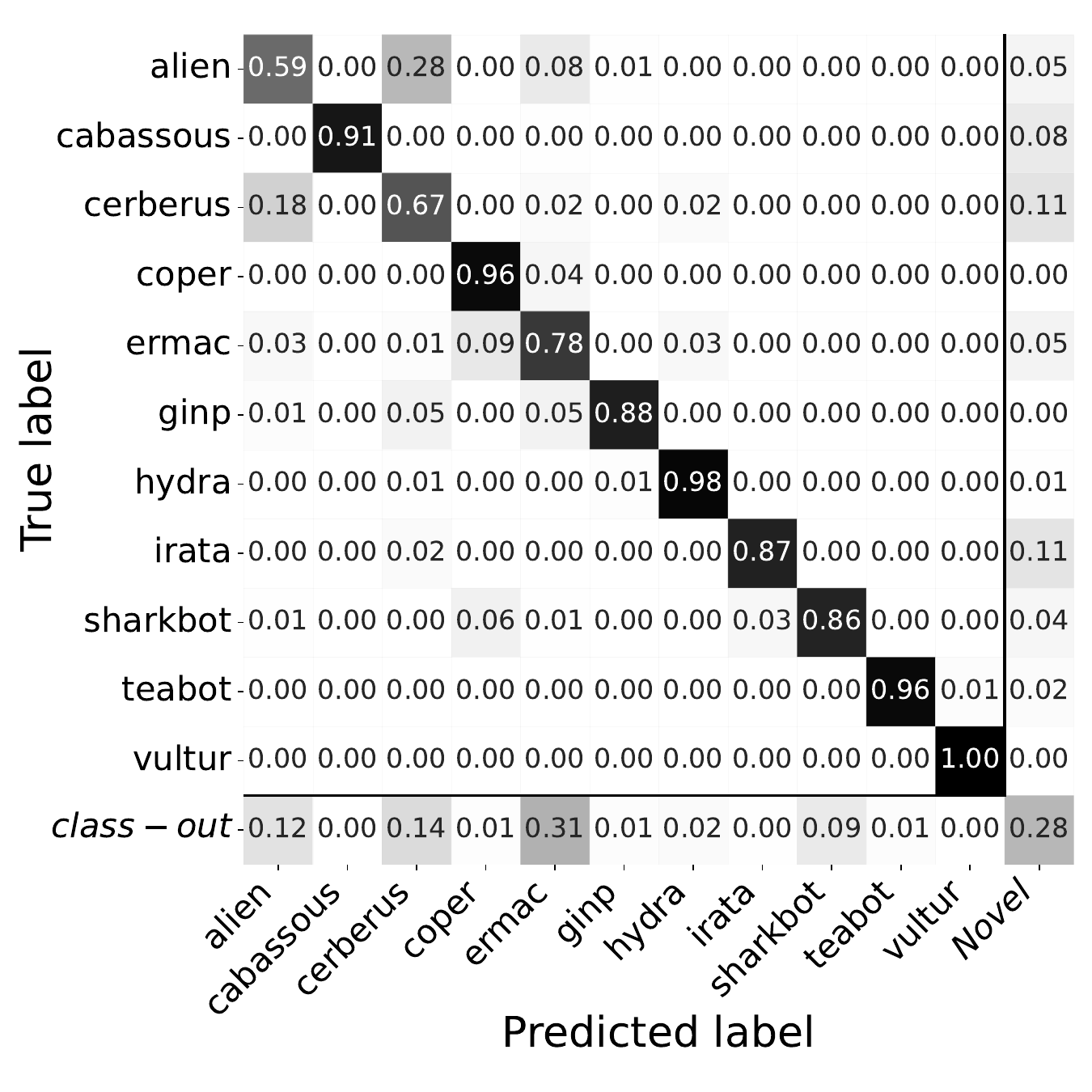}
                    \caption{}
                    \label{fig:loco_kfold_osr_recall_matrix_private}
                \end{subfigure}
                \hfill
                \begin{subfigure}[t]{.28\linewidth}
                    \centering
                    \raisebox{7.7mm}{\includegraphics[width=\linewidth]{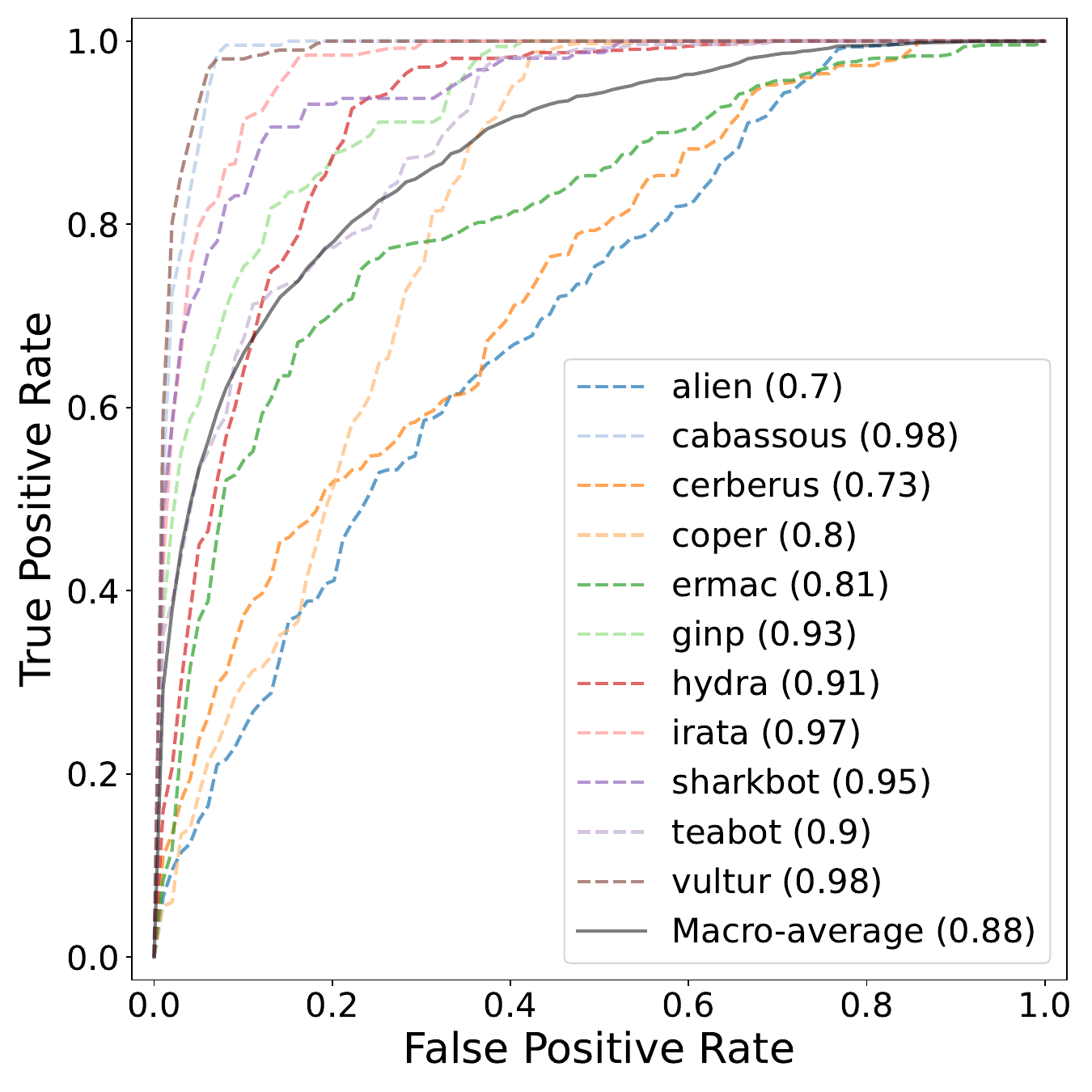}}
                    \caption{}
                    \label{fig:loco_kfold_osr_binary_roc_curves_private}
                \end{subfigure}
                \vspace{0.4cm}
                \caption{\footnotesize Recall confusion matrix of the open-set classifier $\mathcal{K}$, on the public (first row) and proprietary (last row) datasets, when (a),(d) instances of the less populous classes are grouped in the dummy class $others$ and treated as $novel$, and (b),(e) each class is sequentially treated as $novel$ in the leave-one-class-out process, along with the associated novelty detection ROC curves (c),(f).}
                \vspace{0.5cm}
                \label{fig:kfold_loco_kfolr_results_public}
            \end{figure*}
            
            We first assess the effectiveness of the tree-based \gboost classifier $\mathcal{C}$ for closed-set recognition on the $10$-fold experiment on both public and proprietary datasets. A good performance in closed-set recognition is closely tied to the reliability of the subsequent open-set recognition procedure~\cite{VazeHanAl21}. Subsequently, we assess the effectiveness of our open-set recognition classifier $\mathcal{K}$ by comparing it against Open-Set Nearest-Neighbor (\osnn)~\cite{MendesDeSouzaAl17}. In the latter experiment, we further investigate the relation between misclassified samples in the closed-set recognition task and the false positive rate in the open-set recognition task. Ultimately, we discuss the performance of our solution within \cleafy's anti-fraud system since its deployment on their engine.
            
            \medskip
            \subsubsection{Malware classification}
                \cref{tab:kfold_loco_kfold} summarizes the aggregated classification performance of the closed-set classifier $\mathcal{C}$. We observe similar performance levels on the public and proprietary dataset in both \fixnovel and \loonovel scenarios, with slightly better results on the proprietary dataset, and this could be attributed to the smaller number of families to classify ($11$ compared to $54$). For the same reason, when only the top $10$ classes of the public dataset are retained, the micro-average and macro-average recall figures increase. We observe also that dealing with a higher-dimensional space ($1800$ compared to $154$) does not necessarily make it easier to separate different classes, as in the Drebin\textsubscript{10} scenario, $\mathcal{C}$ achieves higher performance despite dealing with a comparable number of classes with respect to the proprietary dataset ($10$ compared to $11$). The slightly lower micro-average recall, compared to the macro-average, is due to some misclassifications within the most populous classes, as the micro-average accounts for different class sizes. Overall, \gboost classifier was considered effective in classifying malware families based on high-dimensional binary permission vectors.

            \medskip
            \subsubsection{Malware family discovery}

                \begin{figure*}[t]
                    \centering
                    \begin{minipage}[t]{.3\textwidth}
                        \centering
                        \raisebox{-1.125cm}{\includegraphics[width=.935\linewidth]{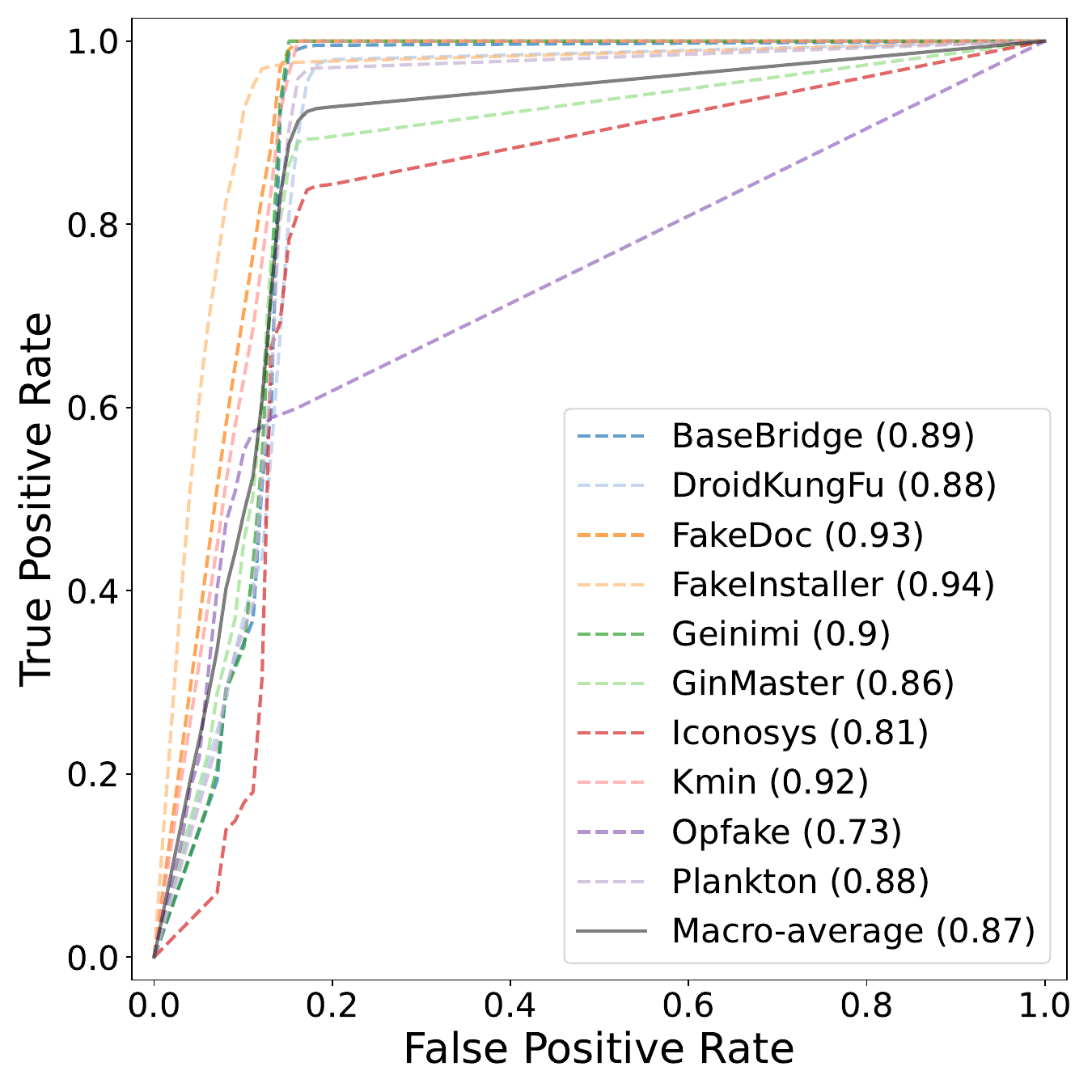}}
                        \vspace{1.825cm}
                        \caption{\footnotesize Novelty detection ROC curves for \osnn on the public dataset, with each class treated as $novel$ in the leave-one-class-out process.}
                        \label{fig:loco_kfold_osr_binary_roc_curves_public_osnn}
                    \end{minipage}
                    \hfill
                    \begin{minipage}{.66\textwidth}
                        \centering
                        \nextfloat
                        \begin{subfigure}[t]{.43\linewidth}
                            \centering
                            \raisebox{11mm}{\includegraphics[width=\linewidth]{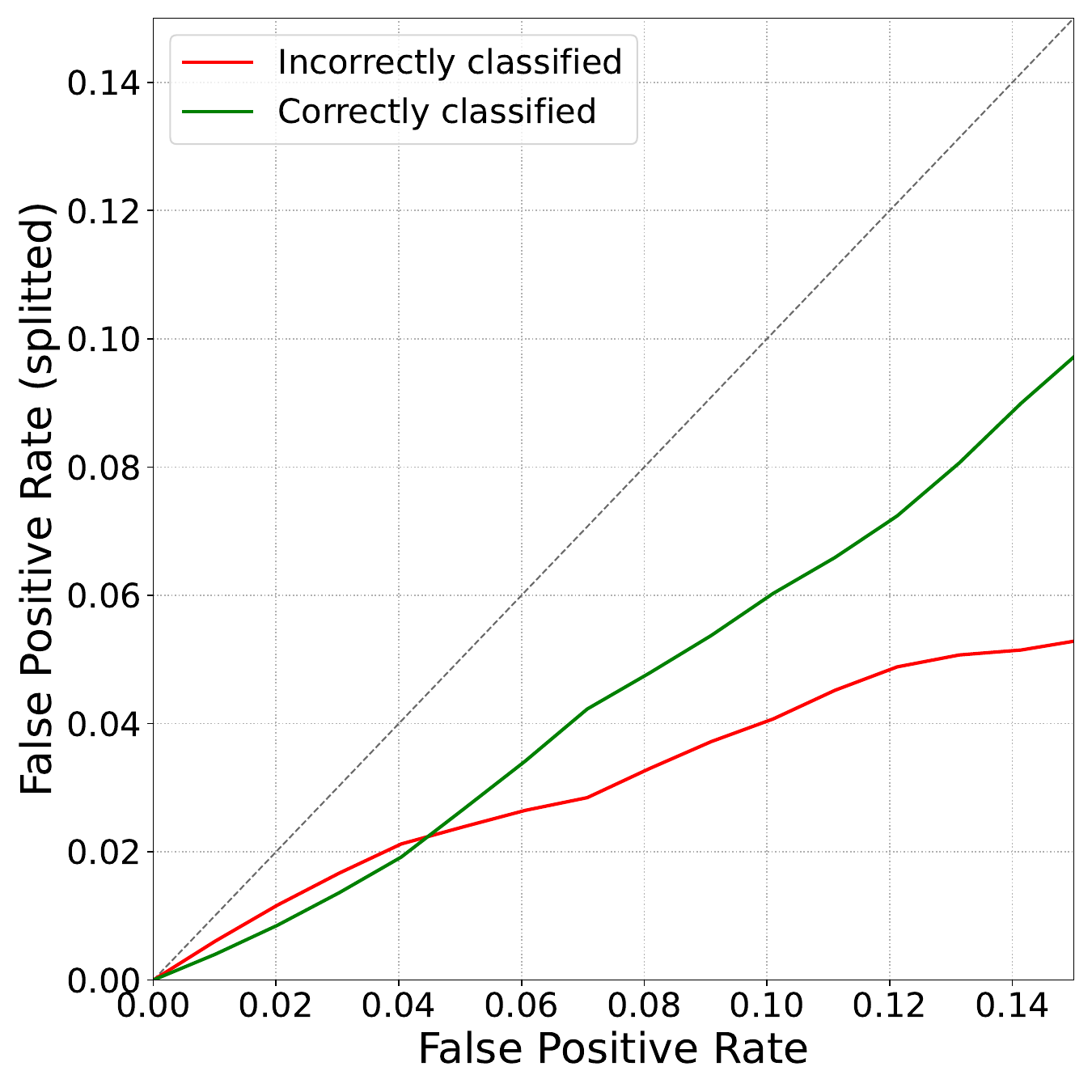}}
                            \caption{}
                            \label{fig:loco_kfold_osr_fpr_decomposition_public}
                        \end{subfigure}
                        \hfill
                        \begin{subfigure}[t]{.54\linewidth}
                            \centering
                            \includegraphics[width=\linewidth]{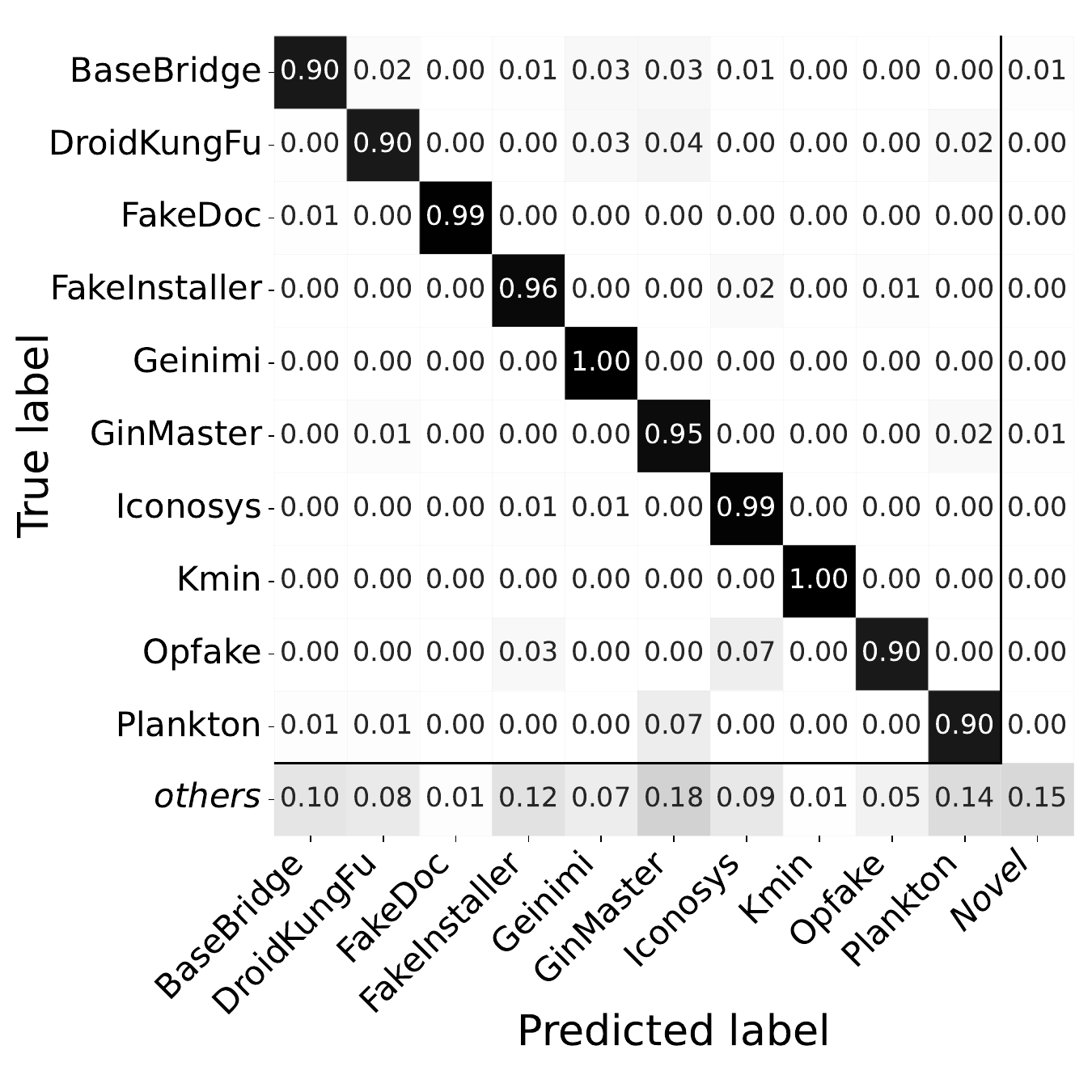}
                            \caption{}
                            \label{fig:loco_kfold_osr_recall_matrix_decomposition_public}
                        \end{subfigure}
                        \vspace{0.4cm}
                        \caption{\footnotesize (a) Different contributions made by samples correctly (green curve) and incorrectly (red curve) classified by $\mathcal{C}$ in composing the total false positive rate of the open-set recognition module $\mathcal{O}$ on the public dataset. (b) Treating the misclassified instances as $novel$ results in better real performance for the open-set classifier $\mathcal{K}$.}
                        \label{fig:loco_kfold_osr_fpr_analysis_public}
                    \end{minipage}
                    \vspace{0.5cm}
                \end{figure*}
            
                \cref{tab:kfold_loco_kfold} also reports the performance of the open-set classifier $\mathcal{K}$, which is, as expected, consistently lower than the closed-set classifier $\mathcal{C}$ due to the additional challenge of recognizing novel malware families and the subsequent impact of false positives.
                
                \cref{fig:kfold_osr_recall_matrix_public,fig:kfold_osr_recall_matrix_private} show the recall confusion matrices of the open-set classifier $\mathcal{K}$ for the \fixnovel scenario, for the public and proprietary datasets respectively. The recall confusion matrices reveal the open-set classifier $\mathcal{K}$ has some false positives, where known malware families are misclassified as $novel$, especially for the proprietary dataset, with $cerberus$ and $irata$ families being the most affected. This suggests that these are classes where the closed-set classifier exhibits great uncertainty, resulting in low confidence levels in classification. We also observe that, while families of the public dataset are easily separable, as shown by the almost clean outer diagonal recall matrix, this does not hold for the proprietary dataset. Specifically, we observe a tendency to misclassify instances belonging to $alien$ and $cerberus$, as evidenced by the prominent errors highlighted. We explain these results as a consequence of the fact that $alien$ is a more recent version of $cerberus$, hence, we expect a substantial overlap in the permissions they require to the system.
                \cref{fig:loco_kfold_osr_recall_matrix_public,fig:loco_kfold_osr_recall_matrix_private} present the recall confusion matrices obtained in the \loonovel scenario, where similar observations to those previously discussed hold.
                \cref{fig:loco_kfold_osr_binary_roc_curves_private,fig:loco_kfold_osr_binary_roc_curves_public} show the ROC curves for the novelty detection task, together with respective AUC (in brackets) when the specified class was considered as the positive $novel$ class. A low AUC value highlights the tendency of the open-set classifier to incorrectly classify samples from unknown families. This happens in the proprietary dataset for the $alien$ and $cerberus$ families due to their similarity, consistent with the classification errors observed before, while in the public dataset, values are overall higher, aligning with the results shown in the confusion matrices.

                \medskip
                We conducted the latter experiment using \osnn as an open-set classifier, and we report the ROC curves and AUC values in~\cref{fig:loco_kfold_osr_binary_roc_curves_public_osnn}. \osnn exhibits very low $TPR$ values at low $FPR$, making it unsuitable for our requirements. In fact, a low $FPR$ is mandatory in our settings, as classifying a sample $\vect{p}$ as belonging to a new family triggers manual inspection, which involves a significant human resource cost. When comparing \osnn to our approach in~\cref{fig:loco_kfold_osr_binary_roc_curves_public}, we observe that at similarly low $FPR$ values, our method achieves significantly higher $TPR$, further highlighting its suitability for the malware family discovery task. The piecewise linear shape of \osnn's ROC curve reflects its limitations in handling high-dimensional binary data. Specifically, \osnn classifies a malware sample as belonging to a new family if the ratio of the distances to its two nearest neighbors from different families falls below a given threshold. However, when dealing with binary data, the number of distinct distance values is limited and highly redundant, leading to a discrete and heavily skewed distribution of distance ratios, and ultimately causing the piecewise progression in $TPR$ values we observe in~\cref{fig:loco_kfold_osr_binary_roc_curves_public_osnn}.
                
                A comparison of computational time between our method and \osnn highlights the superior efficiency of our approach, with inference times of $1.359 \cdot 10^{-5}$ and $1.620 \cdot 10^{-4}$ seconds respectively, showing a difference of more than an order of magnitude. This disparity stems from \osnn’s higher computational complexity, which is dominated by the $O(n \: logn)$ sorting of distances between the query sample and the training dataset.
                
            \medskip
            \subsubsection{False alarms analysis}
                \label{subsubsec:false_alarms_analysis}
                
                In this paragraph, we focus on examining the composition of false alarms on the public dataset Drebin\textsubscript{10}. To this end, we decomposed the false positives of the open-set recognition module into two groups: those originating from instances that were correctly classified by $\mathcal{C}$ before introducing the open-set recognition module $\mathcal{O}$, and those that would have been misclassified regardless of $\mathcal{O}$.
                
                \cref{fig:loco_kfold_osr_fpr_decomposition_public} shows the false positive rate computed over these two groups and we observe that most of the false positives of $\mathcal{O}$, in particular at low $FPR$ values, are samples that would have been misclassified by $\mathcal{C}$ if the open-set module were not in place. This was expected, as instances misclassified by $\mathcal{C}$ are typically associated with low confidence values. We argue that those samples are worth being checked by the threat analysts, since incorrect malware classification results in ineffective solutions against the threat. Therefore, if we do not consider misclassified samples as false positives, and compute the false positive rate only from samples belonging to known families that would have been correctly classified if there was no \osr module $\mathcal{O}$, the false positive rate drops considerably. \cref{fig:loco_kfold_osr_recall_matrix_decomposition_public} displays the corresponding recall confusion matrix when these instances are treated as $novel$, where the model attains a micro-average recall of $0.714$ alongside a macro-average recall of $0.877$.

            \medskip
            \subsubsection{Real-world deployment performance}
                \label{subsubsec:real_world_deployment_performance}
                
                Our solution is currently used by analysts at \cleafy as a complementary tool to enhance malware classification and discover new families. Testing our solution in their operational environment began in the latter half of 2023, and is currently in use in monitoring telemetry gathered by the threat intelligence division, for the analysis of data from worldwide sources.
                
                The closed-set classification performance of our deployed solution resembles closely those reported in experiments on the proprietary dataset, achieving an accuracy of $83\%$ when tested on about $300$ telemetry applications. Unfortunately, the portion of the telemetry data used for testing did not allow us to quantitatively assess the model's performance in detecting new families. The ability to discover new real malware families cannot be assessed on the deployed system by means of strategies like the leave-one-class-out we adopted before, and during our monitoring initiative there have been no cases of completely different malware families within the Android landscape. However, there have been a couple of noteworthy cases. In the first case, the \osr system reported as \textit{novel} a new version of a known malware. In this case, the malware detected as \textit{novel} had a few significant differences with respect to the other, but not enough to consider this to belong to a new family (contrary to $alien$ and $cerberus$). In the second case, a variation of a known malware was labeled as $novel$ due to different permissions requirements compared to the known one. Additionally, a manual review revealed that hundreds of known malware instances classified as $novel$ actually belonged to families not represented in the training set, indicating they were correctly identified as novel. In fact, most of the false alarms concerned malware programs other than banking malware programs (the primary interest for \cleafy's system), and therefore correctly classified as $novel$.
                \Cleafy prefers not to disclose further information on deployment performance for strategic reasons.

    \section{Conclusion and future works}
        \label{sec:concusion_and_future_works}

        In this study, we combined a tree-based \gboost classifier with the \maxlogit open-set recognition technique to tackle the problem of malware family discovery. We conducted comprehensive experiments on both a public and a proprietary dataset to validate the effectiveness of our approach, and discussed its deployment performance in a real-world environment. Furthermore, our analysis on false alarms and the impact of misclassified instances emphasized the practical value of an open-set recognition approach in malware classification.
        
        We envisage several potential future directions. This includes advanced feature engineering techniques to enhance the performance of open-set recognition further. Another interesting direction involves developing a dynamic thresholding system that adjusts the $FPR$ threshold $\tau$ adaptively, taking into account contextual information such as the current threat landscape to achieve the desired sensibility of the \osr module.

    \bibliographystyle{IEEEtran}
    \bibliography{references}
\end{document}